\documentclass[12pt]{article}
 
\usepackage[dvips]{color}                 %%% for colours in the graphs
\usepackage{graphicx}                     %%% for pictures and figures
\usepackage{amssymb}

\usepackage{xspace}                       %%% to get the correct spacing of
\newcommand{\absatz}{\vspace{2ex}\noindent}

\renewcommand{\today}{\ifcase\day\or 1st\or 2nd\or 3rd\or 4th\or 5th\or 6th\or
        7th\or 8th\or 9th\or 10th\or 11th\or 12th\or 13th\or 14th\or 15th\or 
        16th\or 17th\or 18th\or 19th\or 20th\or 21st\or 22nd\or 23rd\or 24th\or
        25th\or 26th\or 27th\or 28th\or 29th\or 30th\or 
        31st\fi~\ifcase\month\or January\or February\or March\or April\or
        May\or June\or July\or August\or September\or October\or November\or
        December\fi \space \number\year}   

\newcommand{\journal}[4]{{#1}\textbf{#2}, #4 (#3)}
\newcommand{\EPJA}{\textit{Eur.\ Phys.\ J.\ }\textbf{A}}

\newcommand{\NPA}{\textit{Nucl.\ Phys.\ }\textbf{A}}
\newcommand{\NPB}{\textit{Nucl.\ Phys.\ }\textbf{B}}
\newcommand{\PLB}{\textit{Phys.\ Lett.\ }\textbf{B}}
\newcommand{\PR}{\textit{Phys.\ Rev.\ }}

\newcommand{\PRC}{\PR\textbf{C}}
\newcommand{\PRD}{\PR\textbf{D}}
\newcommand{\PRL}{\PR\textit{Lett.\ }}

\newcommand{\half}{\frac{1}{2}}

\newcommand{\dd}{\mathrm{d}}

\newcommand{\kv}{\vec{k}}

\newcommand{\mpi}{m_\pi}
\newcommand{\fpi}{f_\pi}
\newcommand{\MeV}{\mathrm{MeV}}

\newcommand{\fm}{\mathrm{fm}}

\newcommand{\HBCPT}{HB$\chi$PT\xspace}

\newcommand{\de}{\partial}
\newcommand{\dev}{\vec{\de}}

\newcommand{\du}{a}

\newcommand{\calO}{\mathcal{O}}

\newcommand{\mytitle}[1]{
                         \begin{center}
                           \LARGE{\textbf{#1}}
                         \end{center}}
\newcommand{\myauthor}[1]{\textbf{#1}}
\newcommand{\myaddress}[1]{\textit{#1}}
\newcommand{\mypreprint}[1]{\begin{flushright} #1 \end{flushright}}

\begin{document}
%%%%%%%%%%%%%%%%%%%%%%%%%%%%%%%%%%%%%%%%%%%%%%%%%%%%%%%%%%%%%%%%%%%%%%%%%%%%%%%
% This is a nice title page including abstract ....
%

\begin{titlepage}
  
  \mypreprint{
    %\textbf{Draft version \today}\hfill
    nucl-th/0110006\\
    TUM-T39-01-22 \\
    ECT*-01-27 \\
    1st October 2001\\
    Re-revised version 13th February 2002,
    \PRC \textbf{65}}
  
  %\vspace*{0.5cm}
  \vspace*{0.1cm}

  \mytitle{Dispersion Effects in Nucleon Polarisabilities}
  
  %\vspace*{0.5cm}
  \vspace*{0.3cm}

\begin{center}
  
  \myauthor{Harald W.\ Grie\3hammer}\footnote{Email:
    hgrie@physik.tu-muenchen.de} and \myauthor{Thomas
    R.~Hemmert}\footnote{Email: themmert@physik.tu-muenchen.de}
  
  \vspace*{0.5cm}
  
  \myaddress{
    Institut f{\"u}r Theoretische Physik (T39), Physik-Department,\\
    Technische Universit{\"a}t M{\"u}nchen, D-85747 Garching,
    Germany\footnote{permanent address}}
  \\[2ex]
  and
  \\[2ex]
  \myaddress{ECT*, Villa Tambosi, I-38050 Villazzano (Trento), Italy}
  
  %\vspace*{0.2cm}

\end{center}

\vspace*{0.5cm}

\begin{abstract}
  We present a formalism to extract the dynamical nucleon polarisabilities
  defined via a multipole expansion of the structure amplitudes in nucleon
  Compton scattering. In contradistinction to the static polarisabilities,
  dynamical polarisabilities gauge the response of the internal degrees of
  freedom of a composed object to an external, real photon field of arbitrary
  energy.  Being energy dependent, they therefore contain additional
  information about dispersive effects induced by internal relaxation
  mechanisms, baryonic resonances and meson production thresholds of the
  nucleon. We give explicit formulae to extract the dynamical electric and
  magnetic dipole as well as quadrupole polarisabilities from low energy
  nucleon Compton scattering up to the one pion production threshold and
  discuss the connection to the definition of static nucleon polarisabilities.
  As a concrete example, we examine the results of leading order Heavy Baryon
  Chiral Perturbation Theory for the four leading spin independent iso-scalar
  polarisabilities of the nucleon.  Finally, we consider the possible r{\^o}le
  of energy dependent effects in low energy extractions of the iso-scalar
  dipole polarisabilities from Compton scattering on the deuteron.
\end{abstract}
\vskip 1.0cm
\noindent
\begin{tabular}{rl}
Suggested PACS numbers:& 14.20.Dh, 13.40.-f, 13.60.Fz\\[1ex]
Suggested Keywords: &\begin{minipage}[t]{11cm}
                    Effective Field Theory, nucleon polarisabilities,\\
                    Compton scattering
                    \end{minipage}
\end{tabular}

\vskip 1.0cm

\end{titlepage}

\setcounter{page}{2} \setcounter{footnote}{0} \newpage
  
%%%%%%%%%%%%%%%%%%%%%%%%%%%%%%%%%%%%%%%%%%%%%%%%%%%%%%%%%%%%%%%%%%%%%%%%%%%%%%%
%%%%%%%%%%%%%%%%%%%%%%%%%%%%%%%%%%%%%%%%%%%%%%%%%%%%%%%%%%%%%%%%%%%%%%%%%%%%%%%
%%%%%%%%%%%%%%%%%%%%%%%%%%%%%%%%%%%%%%%%%%%%%%%%%%%%%%%%%%%%%%%%%%%%%%%%%%%%%%%
% Main Body
%

%%%%%%%%%%%%%%% Intro %%%%%%%%%%%%%%%%%%%
\section{Introduction}
\setcounter{equation}{0}
\label{sec:intro}
%%%%%%%%%%%%%%%%%%%%%%%%%%%%

The study of the low energy structure of the nucleon via Compton scattering
$\gamma N \rightarrow \gamma^\prime N$ has a long history. For a proton
target, one is tempted to obtain a good description of the resulting cross
section up to photon energies of about $50\;\MeV$ by assuming that one
scatters off a positively charged point-like spin 1/2 particle with an
additional (Pauli) anomalous magnetic moment of $\kappa_p=1.79$ nuclear
magnetons. For higher photon energies, this simple picture breaks down as more
details of the complicated internal structure of the nucleon become visible to
the electro-magnetic probe. Even at low energies the description can be
improved by noting that there are additional effects due to the target
structure which turn out to be nearly as large as the magnetic moment terms.
It has been argued (see e.g.~\cite{babusci}) that to first order one can take
these effects into account via an effective Hamiltonian ansatz
\begin{eqnarray}
  \label{eq:hamiltonian1}
  H_\mathrm{eff}=
  -\frac{1}{2}\;4\pi\left(\bar{\alpha}_E\;\vec{E}^2+\bar{\beta}_M\;\vec{B}^2
  \right)\;\;,
\end{eqnarray}
where $\bar{\alpha}_E\; (\bar{\beta}_M)$ constitutes the electric (magnetic)
polarisability of the nucleon in response to an external electric (magnetic)
field $\vec{E}\;(\vec{B})$ generated by the incoming and outgoing photon. In
the past few decades, a lot of effort has been invested to determine these
fundamental structure parameters of the nucleon.  The most recent global
analysis of the world data for the proton \cite{Olmos} gives
\begin{eqnarray}
          \label{eq:globala}
   \bar{\alpha}_p&=&\left(11.9\pm0.5\mp0.5\right)\times10^{-4}\;\fm^3
   \nonumber\\
   \bar{\beta}_p&=&\left(1.5\pm0.6\pm0.2\right)\times 10^{-4}\;\fm^3\;\;, 
\end{eqnarray}
whereas the neutron polarisabilities are less well determined, with sometimes
conflicting experiments, see e.g.~recently \cite{Kolb00,Levchuk01} and
references in~\cite{hggr}
\begin{eqnarray}
  \label{eq:globalb}
   \bar{\alpha}_n&=&[0\dots16.6]\times 10^{-4}\;\fm^3\nonumber\\
   \bar{\beta}_n&=&[1.2\dots14]\times 10^{-4}\;\fm^3\;\;. 
\end{eqnarray}
Here, we do not want to comment on the accuracy, error bars or inherent
problems of the neutron measurements. Rather, we want to point out that the
numbers displayed in (\ref{eq:globala}/\ref{eq:globalb}) attempt to describe
the nucleon's response to a \emph{static} external electro-magnetic
$\vec{E},\,\vec{B}$ field, which in the language of nucleon Compton scattering
corresponds to the limit that the energy of the incoming photon goes to zero.
Obviously, the experiments from which the numbers of (\ref{eq:globala}) have
been derived were not performed at zero energy as there is hardly any
sensitivity to the polarisabilities in this kinematic regime.  In fact, most
of the recent proton Compton scattering experiments were performed with $55$
to $800\;\MeV$ photons \cite{Olmos,Galler,Blanpied} and thus had to rely on
additional, theoretical input to relate the results to zero energy parameters.
In other words, one needs to know the \emph{energy dependence} of the various
nucleon structure parameters in order to extract static nucleon properties.
For photon energies below the two pion production threshold, the most
convenient way to take these effects into account is the use of dispersion
theory.  Its starting point is a dispersion relation for each of the six
invariant amplitudes governing spin 1/2 Compton scattering.  The static
polarisabilities defined at zero energy can then be related to a subtraction
constant of such a (subtracted) dispersion relation, whereas the energy
dependent effects are subsumed into the integral over the photo-absorption
cross sections $\gamma N\rightarrow X$. Given that there is enough
experimental information on the required inelastic cross sections, one can
then calculate differential cross sections for nucleon Compton scattering and
finally fit the unknown subtraction constants (namely the polarisabilities) to
reproduce the measured Compton cross sections. Details on the use of
subtracted and unsubtracted dispersion relations in the analysis of medium
energy nucleon Compton scattering can be found in~\cite{lvov,drechsel}.  Here,
we only want to remind the reader that even in the usual extraction of the
static polarisabilities, the correct treatment of the energy dependent effects
is crucial. In the following, we argue that this energy dependence is not just
a dirty side-effect to be subsumed into a dispersion integral, but that the
effect is interesting in its own right, leading to a better understanding of
the degrees of freedom which rule the electric and magnetic properties of the
nucleon\footnote{In addition to electric and magnetic polarisabilities, a spin
  1/2 nucleon possesses so called spin polarisabilities~\cite{spin} which can
  also be described via an effective Hamiltonian ansatz but do not have a
  direct analogue in classical electrodynamics. The method and line of
  thinking developed here can also be extended to a dynamical description and
  interpretation of these fundamental low energy nucleon spin structure
  parameters~\cite{hgth}, but this is beyond the scope of this presentation.}.

\absatz The article is organised as follows: In the next section, we discuss a
multipole expansion for the nucleon amplitudes of Compton scattering which
also serves as point of reference for a discussion of the energy dependent
effects expected in nucleons at the end of the same section.
Section~\ref{sec:matching} shows how to extract the dynamical polarisabilities
from given Compton amplitudes. We then exemplify our method by calculating the
leading one loop order Heavy Baryon Chiral Perturbation prediction for the
electric and magnetic dipole and quadrupole polarisabilities
(Sect.~\ref{sec:example}).  Before concluding by summarising and discussing
our findings, we give an outlook on possible implications of our results on
the extraction of nucleon polarisabilities from Compton scattering off the
deuteron in Sect.~\ref{sec:comments}.

%%%%%%%%%%%%%%% Intro %%%%%%%%%%%%%%%%%%%
\section{Multipole Expansion}
\setcounter{equation}{0}
\label{sec:multipole}
%%%%%%%%%%%%%%%%%%%%%%%%%%%%

The $T$ matrix of nucleon Compton scattering can be written in terms of six
amplitudes $A_i(\omega,z),\;i=1,\dots,6$. In the following, we will work in
the centre-of-mass (cm) frame and use the Compton amplitudes of~\cite{spin}.
Thus, $\omega$ denotes the cm energy of a real photon scattering under the cm
angle $\theta$ off the nucleon, with $z=\cos\theta$.  Polarisabilities are
designed to be a measure of the excitation spectrum of a system, but low
energy Compton scattering off the proton is dominated by Born terms, i.e.~by
the successive interactions of two photons with a point-like nucleon with
charge Q and anomalous magnetic moment $\kappa$. It is therefore customary to
subtract these ``nucleon pole'' effects from the six Compton amplitudes and
only analyse the remainder in terms of polarisabilities. Clearly, the
differential cross sections are independent of this artificial separation of
the amplitudes into ``pole'' and ``non-pole'' parts.  Here, we only state the
principle
\begin{eqnarray}
   \bar{A}_i(\omega,z)&=&
   A_i(\omega,z)-A_i^\mathrm{pole}(\omega,z),\;i=1,\dots,6
   \label{eq:Born}
\end{eqnarray}
with the caveat that any theoretical calculation of nucleon polarisabilities
should state carefully which definition of ``nucleon pole'' terms, which set
of Compton amplitudes and -- in the case of non-relativistic frameworks --
which frame of reference was used. A covariant definition of ``nucleon pole''
terms can be found e.g.~in~\cite{babusci}. In the following, we will only
discuss the ``non-pole'' amplitudes $\bar{A}_i$.

The electric and magnetic polarisabilities of the nucleon are contained in the
two \emph{spin independent} structure amplitudes $\bar{A}_1,\;\bar{A}_2$:
\begin{eqnarray}
  \label{eq:comptonstructure}
   \bar{T}(\omega,z)=&\bar{A}_1(\omega,z)\;
   \vec{\epsilon}^{\prime\,\ast}\cdot\vec{\epsilon}+
   \bar{A}_2(\omega,z)\;
   \left(\vec{\epsilon}^{\prime\,\ast}\cdot\hat{\kv}\right)\,
   \left(\vec{\epsilon}\cdot\hat{\kv}{}^\prime\right)+\ldots\;\;,
\end{eqnarray}
where $\hat{\kv}=\kv/\omega$ ($\hat{\kv}{}^\prime=\kv^\prime/\omega$) is the
unit vector in the direction of the momentum of the incoming (outgoing)
photon.

A multipole expansion for Compton scattering has been defined a long time ago
\cite{Ritus,Ritus2,Ritus3} in a different basis $R_i$. With the connecting
formula
\begin{eqnarray}
  \label{eq:connecting}
  \bar{A}_1(\omega,z)&=&\frac{4\pi\,W}{M}\left[\bar{R}_1(\omega,z)+
                        z\,\bar{R}_2(\omega,z)\right]\nonumber\\
  \bar{A}_2(\omega,z)&=&-\,\frac{4\pi\,W}{M}\,\bar{R}_2(\omega,z)\;\;,
\end{eqnarray}
where $W=\omega+\sqrt{M^2+\omega^2}$ is the total cm energy and $M$ the
nucleon mass, we can easily utilise the known multipole expansion of the
structure dependent (``non-pole'') part $\bar{R}_1,\,\bar{R}_2$ of the
functions $R_1,\,R_2$ \cite{Ritus,Ritus2,Ritus3}:
\begin{eqnarray}
  \label{eq:multipoleexpansion}
  \bar{R}_1(\omega,z)&=&
      \sum\limits_{l=1}^{\infty}\Big\{\Big[(l+1)f_{EE}^{l+}(\omega)+
      l\,f_{EE}^{l-}(\omega)\Big]\Big(l\,P_l^\prime(z)+
      P_{l-1}^{\prime\prime}(z)\Big)-\nonumber\\
      &&\phantom{\sum\limits_{l=1}^{\infty}\Big\{}
      -\Big[(l+1)f_{MM}^{l+}(\omega)+l\,f_{MM}^{l-}(\omega)\Big]
      P_l^{\prime\prime}(z)\Big\} \nonumber\\
  \bar{R}_2(\omega,z)&=&
      \sum\limits_{l=1}^{\infty}\Big\{\Big[(l+1)f_{MM}^{l+}(\omega)+
      l\,f_{MM}^{l-}(\omega)\Big]\Big(l\,P_l^\prime(z)+
      P_{l-1}^{\prime\prime}(z)\Big)-\nonumber\\
      &&\phantom{\sum\limits_{l=1}^{\infty}\Big\{}
      -\Big[(l+1)f_{EE}^{l+}(\omega)+l\,f_{EE}^{l-}(\omega)\Big]
      P_l^{\prime\prime}(z)\Big\}\;\;,
\end{eqnarray}
where $P_i^{(n)}$ denotes the $n$th derivative of the Legendre polynomial
$P_i(z)$ with respect to $z$. The multipole amplitudes
$f_{TT^\prime}^{l\pm}(\omega)$ with $T,T^\prime=E,M$ contain the energy
dependence and correspond to transitions $T l\rightarrow T^\prime l^\prime$,
where $l$ corresponds to the angular momentum of the initial photon and
$l^\prime=l\pm\{1,\,0\}$ to the one of the final photon. The total angular
momentum is $l\pm=j=l\pm\half$. We now \emph{define}\footnote{In contrast
  to~\cite{babusci}, we take (\ref{eq:polarisabilitiesdef}) as an exact
  definition of the dynamical polarisabilities in the cm frame. We therefore
  do not require the polarisabilities thus defined to be even functions of
  $\omega$. Factorising out the purely kinematical term $W/M$ in
  (\ref{eq:connecting}) removes a trivial energy dependence of the
  polarisabilities which varies with the nucleon mass.}  energy dependent
electric and magnetic polarisabilities via the relations
\begin{eqnarray}
  \label{eq:polarisabilitiesdef}
   \alpha_{El}(\omega)&=&N_l\,
                      \frac{l\left(l+1\right)\left[\left(l+1\right)
                      f_{EE}^{l+}+l\,f_{EE}^{l-}\right]}{2\,\omega^{2l}}
                  \nonumber\\
   \beta_{Ml}(\omega)&=&N_l\,
                     \frac{l\left(l+1\right)\left[\left(l+1\right)
                     f_{MM}^{l+}+l\,f_{MM}^{l-}\right]}{2\,\omega^{2l}}
\end{eqnarray}
with $N_1= 1,\;N_2=12,\;N_l=\frac{2l}{l+1}\,[(2l-1)!!]^2$~being normalisation
factors~\cite{Gui} chosen to recover the usual static polarisabilities in the
limit of zero photon energy $\omega\rightarrow 0$. For example, the well known
static electric and magnetic polarisabilities $\bar{\alpha}_E,\;
\bar{\beta}_M$ of the nucleon are recovered as the low energy limit of two
electric or magnetic dipole transitions
\begin{eqnarray}
  \bar{\alpha}_E=\lim_{\omega\rightarrow 0}\alpha_{E1}(\omega)\;\;& &
  \bar{\beta}_M=\lim_{\omega\rightarrow 0}\beta_{M1}(\omega)\;\;,
\end{eqnarray}
leading to an excitation of the nucleon and a successive de-excitation with
the same multipole character\footnote{A mixing of multipolarities between
  excitation and de-excitation only occurs for spin polarisabilities.}.
However, as (\ref{eq:polarisabilitiesdef}) indicates, there is no need to end
the discussion of nucleon structure as tested in Compton scattering at the
lowest order of the non-pole level, where only dipole polarisabilities are
tested. For example, the static electric and magnetic quadrupole
polarisabilities $\bar{\alpha}_{E2},\;\bar{\beta}_{M2}$ originating from the
effective Hamiltonian (see e.g.~\cite{babusci})
\begin{eqnarray}
  \label{eq:hamiltonian2}
  H_\mathrm{eff}&=&-\frac{4\pi}{12}\;\Big[\bar{\alpha}_{E2}\;E_{ij}^2\;+
   \;\bar{\beta}_{M2}\;B_{ij}^2\Big]\;\;,
\end{eqnarray}
with $T_{ij}:=\half (\de_i T_j + \de_j T_i),\;T=E,B$, are recovered via
\begin{eqnarray}
  \bar{\alpha}_{E2}=\lim_{\omega\rightarrow 0}\alpha_{E2}(\omega)\;\;& &
  \bar{\beta}_{M2}=\lim_{\omega\rightarrow 0}\beta_{M2}(\omega)\;\;.
\end{eqnarray}
Higher order (static and dynamical) multipoles can be defined at will.

\absatz While we follow the definitions of Ref.~\cite{babusci} for the static
polarisabilities, we explicitly keep the energy dependence of the dynamical
polarisabilities as given via (\ref{eq:polarisabilitiesdef}). This corresponds
to defining the coefficients $\alpha_{El},\;\beta_{Bl}$ in the Hamiltonians
(\ref{eq:hamiltonian1}/\ref{eq:hamiltonian2}) as energy dependent, as we
demonstrate now.

%\footnote{
In the spin independent sector, we can absorb the polarisability-like
interactions containing time derivatives of the electric or magnetic field as
defined in~\cite{babusci} into the definitions of the \emph{dynamical}
polarisabilities, e.g.
\begin{eqnarray}
  \label{eq:absorb}
  -\frac{4\pi}{2}\;\Big[\bar{\alpha}_E\;\vec{E}^2+
  \bar{\alpha}_{E\nu}\;\dot{\vec{E}^2}+\dots\Big]\;\to\;
  -\frac{4\pi}{2}\;\alpha_{E1}(\omega)\;\vec{E}^2\;\;.
\end{eqnarray}
At small enough photon energies, this definition is equivalent to the Taylor
expansion of~\cite{babusci},
\begin{eqnarray}
  \label{eq:taylor}
  \lim\limits_{\omega\to0}\alpha_{E1}(\omega)=\bar{\alpha}_E+
  \bar{\alpha}_{E\nu}\;\omega^2+\dots\;\;,
\end{eqnarray}
which clearly has the dis-advantage that the series does not necessarily
converge at higher energies. The effective Hamiltonian of the dynamical
polarisabilities\footnote{Terms containing $\dev\times\vec{E}$ or
  $\dev\times\vec{B}$ are absent in this Hamiltonian after using Maxwell's
  equations to convert them into time derivatives of $\vec{B}$ or $\vec{E}$.}
\begin{equation}
  \label{eq:ourhamiltonian}
  H_\mathrm{eff}^\mathrm{dyn}=
  -4\pi\;\Big[\half\;\Big(\alpha_{E1}(\omega)\;\vec{E}^2\;+
   \;\beta_{M1}(\omega)\;\vec{B}^2\Big) 
   +\frac{1}{12}\;\Big(\alpha_{E2}(\omega)\;E_{ij}^2\;+
   \;\beta_{M2}(\omega)\;B_{ij}^2\Big)\Big]\;+\;\dots
\end{equation}
is thus built only of the spatial multipoles of the photon field, with all
temporal response and retardation effects contained in the definition of the
dynamical polarisabilities.
%Notice that in such a formulation, the kinematical pre-factor $W/M$ in the
%Compton amplitudes (\ref{eq:connecting}) is produced order by order in
%perturbation theory.
%}.

\absatz We believe that the study of this energy dependence is interesting in
its own right as the temporal response of the nucleon to external
electro-magnetic perturbations is encoded therein. Recall that
polarisabilities are a \emph{global}\footnote{So called ``generalised
  polarisabilities'' defined in \cite{Guichon,L'vov:2001fz} can be extracted
  from analyses of virtual Compton scattering experiments $\gamma^\ast
  N\rightarrow\gamma N$ and provide \emph{local} information on the internal
  nucleon dynamics.} measure of the low energy excitation spectrum of a
system, and hence mirror the response of the internal degrees of freedom to an
external, real photon field. A composite object will therefore necessarily
have energy dependent polarisabilities.  We also notice that the concept of
dispersive effects for the polarisabilities of composite systems is textbook
knowledge in classical electro-dynamics and solid state physics. It was also
utilised in the Sixties in nuclear structure, mostly in the context of giant
dipole resonances of heavy nuclei, see e.g.~\cite{sixties} and references
therein.

It is well known from many branches of physics that polarisabilities can
become energy dependent due to internal relaxation mechanisms, resonances and
particle production thresholds in a physical system.  In the first, some of
the internal degrees of freedom of the system ``freeze out'' at some scale as
the photon energy is increased because they cannot respond any more to the
rapidly varying photon fields. This leads to a drop of the polarisabilities
stemming from them as their relaxation time becomes larger than the inverse
frequency of the photon field.  Resonance effects are traditionally discussed
in the Lorentz model which tries to describe their influence on
polarisabilities by assuming that the photon field couples to a dampened
harmonic multipole oscillator with a proper frequency given by the resonance
energy. The width over which the polarisability changes and its maximum are
then related to the decay width and strength of the resonance. In the nucleon,
we therefore expect explicit $\Delta$, $N^\ast$ excitations etc.~to make the
polarisabilities energy dependent.  Finally, production thresholds modify the
energy dependence of the polarisabilities, adding above threshold to the
dynamical polarisabilities imaginary parts which are necessary to describe the
break-up channels not contained in the polarisability Hamiltonian
(\ref{eq:ourhamiltonian}). The scale over which the polarisabilities vary is
in this case set by the mass of the particle produced, and by the strength of
the threshold. In the nucleon, the one pion production threshold is the most
prominent example.  Albeit qualitative in nature, this discussion demonstrates
which important information the energy dependence of the polarisabilities
contains on the internal degrees of freedom of a system.

We hope to have excited the reader's imagination about the physics
possibilities of studying the energy dependence in dynamical polarisabilities
defined via a multipole expansion. We now move on to simple formulae which
allow to extract the energy dependence of the leading four spin independent
polarisabilities discussed above from any model of nucleon Compton scattering.

%%%%%%%%%%%%%%%%%%%%%%%%%%%%
\section{Matching Formulae}  
\setcounter{equation}{0}
\label{sec:matching}
%%%%%%%%%%%%%%%%%%%%%%%%%%%%

Knowledge of the energy and angular dependence of the pole subtracted
structure amplitudes $\bar{A}_1(\omega,z)$ and $ \bar{A}_2(\omega,z)$ in the
cm frame as defined in (\ref{eq:comptonstructure}) suffices to extract the
dynamical electric and magnetic dipole and quadrupole polarisabilities of the
nucleon. Given these two functions, one has to decide at which angular
momentum $l$ to truncate the multipole expansion. Although no requirement of
the multipole expansion, we choose to work at energies not higher than the one
pion production threshold in Compton scattering to keep the number of
multipole polarisabilities small. It is trivial that the higher the photon
energy, the more of the higher multipoles have to be included to obtain an
accurate result. For any given calculation, it is also simple to check how
many multipole polarisabilities defined via (\ref{eq:polarisabilitiesdef})
have to be included in the expansion until the polarisabilities of interest
are stable e.g.~at the $5\%$ level. Below, we focus on the first four,
$\alpha_{E1}(\omega),\,\beta_{M1}(\omega),\,\alpha_{E2}(\omega)$ and
$\beta_{M2}(\omega)$.

The spin independent Compton structure amplitudes $\bar{A}_1,\;\bar{A}_2$ in
the cm frame, truncated at $l=3$, read using (\ref{eq:connecting}) to
(\ref{eq:polarisabilitiesdef})
\begin{eqnarray}
  \label{eq:amplitudes1}
  \bar{A}_1(\omega,z)\Big|_{l=3}&=&\frac{4\pi\,W}{M}\Bigg[
  \Big(\alpha_{E1}(\omega)+z\beta_{M1}(\omega)\Big)\omega^2+
  \frac{1}{12}\Big(z\alpha_{E2}(\omega)+(2 z^2-1)\beta_{M2}(\omega)
  \Big)\omega^4+
  \nonumber\\
  &&+\;\frac{1}{4N_3}\Big( (5z^2-1)\alpha_{E3}(\omega)+
     z(15z^2-11)\beta_{M3}(\omega)\Big)\omega^6\Bigg]
     \nonumber\\
  \bar{A}_2(\omega,z)\Big|_{l=3}&=&-\;\frac{4\pi\,W}{M}
  \Bigg[
  \beta_{M1}(\omega)\;\omega^2+
  \frac{1}{12}\Big(-\alpha_{E2}(\omega)+2 z\beta_{M2}(\omega)
  \Big)\omega^4+
  \nonumber\\
  &&+\;\frac{1}{4N_3}\Big(-10z \alpha_{E3}(\omega)+
     (15z^2-1)\beta_{M3}(\omega)\Big)\omega^6\Bigg]\;\;.
\end{eqnarray}
Systematically eliminating all dependence on the octupole polarisabilities
$\alpha_{E3},\,\beta_{M3}$ and their normalisation factors $N_3$ contained in
(\ref{eq:amplitudes1}), the system of equations for the four leading electric
and magnetic dipole and quadrupole polarisabilities in the cm system is easily
solved:
\begin{eqnarray}
  \label{eq:polarisabilitiesextr1}
  \alpha_{E1}(\omega)\Big|_{l=3}&=&\frac{M}{4\pi\,W\,\omega^2}\;\Big(
                \bar{A}_1(\omega,z)\Big|_{z=0}-
                \frac{1}{10}\;\bar{A}_2^\prime(\omega,z)\Big|_{z=0}
                +\frac{1}{5}\;\bar{A}_1^{\prime\prime}(\omega,z)\Big|_{z=0}
                \Big)
                \nonumber\\
  \beta_{M1}(\omega)\Big|_{l=3}&=&\frac{M}{8\pi\,W\,\omega^2}\;
            \Big(\bar{A}_1^\prime(\omega,z)\Big|_{z=0}-
                \bar{A}_2(\omega,z)\Big|_{z=0}-\frac{2}{5}\,
                \bar{A}_2^{\prime\prime}(\omega,z)\Big|_{z=0}
                \Big)\nonumber\\
  \alpha_{E2}(\omega)\Big|_{l=3}&=&   \frac{3M}{2\pi\,W\,\omega^4}\;
            \Big(\bar{A}_1^\prime(\omega,z)\Big|_{z=0}+
                \bar{A}_2(\omega,z)\Big|_{z=0}-\frac{1}{3}\,
                \bar{A}_2^{\prime\prime}(\omega,z)\Big|_{z=0}
                \Big)
                \nonumber\\
  \beta_{M2}(\omega)\Big|_{l=3}&=&\frac{M}{2\pi\,W\,\omega^4}\;
            \Big(\bar{A}_1^{\prime\prime}(\omega,z)\Big|_{z=0}-
                \bar{A}_2^{\prime}(\omega,z)\Big|_{z=0}\Big)\;,
\end{eqnarray}
The prime denotes again differentiation with respect to $z=\cos\theta$ in the
cm system.

The formulae given in~(\ref{eq:polarisabilitiesextr1}) constitute the central
result of our investigation. They are valid up to the influence of
electro-magnetic $l\ge4$ transitions\footnote{As the system is over-determined,
  there exist alternative expressions for $\beta_{M1}(\omega)$ and
  $\alpha_{E2}(\omega)$,
\begin{eqnarray}
  \label{eq:polarisabilitiesextr2}
  \beta_{M1}(\omega)\Big|_{l=3}&=&\frac{M}{8\pi\,W\,\omega^2}\;
            \Big(\bar{A}_1^\prime(\omega,z)\Big|_{z=0}-
                \bar{A}_2(\omega,z)\Big|_{z=0}+\frac{2}{15}\,
                \bar{A}_1^{\prime\prime\prime}(\omega,z)\Big|_{z=0}
                \Big)\nonumber\\
  \alpha_{E2}(\omega)\Big|_{l=3}&=&   \frac{3M}{2\pi\,W\,\omega^4}\;
            \Big(\bar{A}_1^\prime(\omega,z)\Big|_{z=0}+
                \bar{A}_2(\omega,z)\Big|_{z=0}+\frac{1}{9}\,
                \bar{A}_1^{\prime\prime\prime}(\omega,z)\Big|_{z=0}
                \Big)\;,                
\end{eqnarray}
which differ from the forms given in (\ref{eq:polarisabilitiesextr1}) by
contributions from $l\ge4$ and contain third derivatives. The difference
between both definitions can thus serve for each of the afflicted multipoles
as a measure of the validity of the $l=3$ truncation in this energy range. If
one is interested in dynamical polarisabilities beyond the quadrupole level,
this finding can easily be generalised: Truncating at multipolarity $l$, there
are two ways to extract the highest magnetic multipole $\beta_{Ml}$, and this
affects also the lower multipoles $\beta_{M(l-2n)}$, $\alpha_{E(l-2n+1)}$,
$n=0,\dots,l/2$.}  and can be applied to any theoretical framework that
provides information on the spin independent Compton structure amplitudes
$\bar{A}_1,\,\bar{A}_2$. In the example of the iso-scalar nucleon
polarisabilities from leading order Heavy Baryon Chiral Perturbation Theory
(\HBCPT) discussed in the next section, we will find that effects from $l\ge4$
can safely be neglected and the formulae given in
(\ref{eq:polarisabilitiesextr1}) are by far sufficient to analyse dynamical
dipole and quadrupole polarisabilities for photon energies up to the one pion
production threshold.  We will use this example to demonstrate the feasibility
of the kind of studies proposed here. The validity and accuracy of the
truncation has of course to be checked separately for each theoretical
framework one wants to compare to.  This might necessitate to go to $l=4$ or
higher for some specific model.

We stress again that the multipole expansion of the Compton amplitudes as
performed in (\ref{eq:amplitudes1}) is independent of the decision which of
the polarisabilities are to be extracted in (\ref{eq:polarisabilitiesextr1}).
Of course, wanting to extract the $l_0$th multipole polarisabilities, one has
to expand the amplitude at least to the $l_0$th multipole. However, in that
case the $l_0$th multipole polarisabilities extracted will also be most
sensitive to effects from all the neglected higher multipole polarisabilities
$l>l_0$. It is therefore prudent to multipole expand (\ref{eq:amplitudes1}) to
high multipoles $l>l_0$, and then limit one in solving the resulting system of
equations for the polarisabilities to the first few, $l_0$ multipole
polarisabilities.  As long as the polarisabilities corresponding to higher
multipoles are getting smaller and smaller, effects from higher multipoles
diminish as $l$ increases.

Finally, we remark that in principle one could also derive our result
(\ref{eq:polarisabilitiesextr1}) by utilising projection operators that allow
the extraction of individual Compton multipoles $f_{TT'}^{l\pm}$ from a set of
Compton helicity amplitudes as given by Pfeil et al.~\cite{Pfeil} and then
proceed to reconstruct our definition of dynamical polarisabilities via
(\ref{eq:polarisabilitiesdef}).  As we wish to emphasise idea and physical
content of the dynamical polarisabilities in this article, we postpone this
more formal side to a future publication~\cite{hgth}. Here, it suffices to
remark that in the case of the \HBCPT results presented in the next section,
we see no change in the quadrupole and dipole polarisabilities when going from
$l=2$ to $l=3$ in the expansion of the amplitudes (\ref{eq:amplitudes1}).
Therefore, the $l=3$ result is in that case for all practical purposes as good
as the one obtained from a mathematically rigorous projection formalism.

%%%%%%%%%%%%%%% Intro %%%%%%%%%%%%%%%%%%%
\section{Iso-scalar Nucleon Polarisabilities in Leading Order Heavy Baryon
  Chiral Perturbation Theory} \setcounter{equation}{0}
\label{sec:example}
%%%%%%%%%%%%%%%%%%%%%%%%%%%%

We now turn as a simple example to the prediction \HBCPT gives at leading one
loop order for the dynamical polarisabilities in Compton scattering below the
one pion production threshold. We will see that our master formula
(\ref{eq:polarisabilitiesextr1}) leads to numerically stable results for the
dipole and quadrupole multipoles.

\begin{figure}[!htb]
  \begin{center}
    \includegraphics*[width=0.9\textwidth]{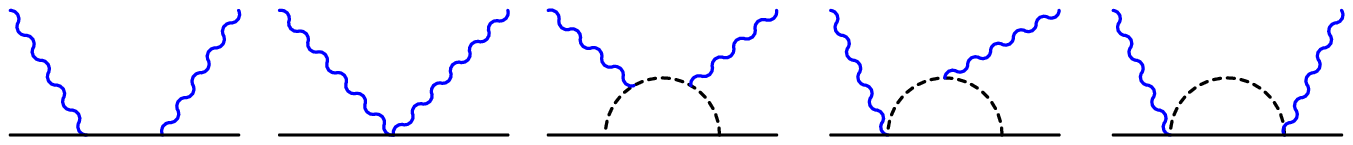}
    \caption{The Compton amplitudes $A_1$ and $A_2$ at leading one loop order 
      in \HBCPT. Graphs obtained by permuting vertices or external lines are
      not displayed.}
    \label{fig:comptonamplitude}
  \end{center}
\end{figure}

The diagrams contributing to the iso-scalar Compton amplitudes are listed in
Fig.~\ref{fig:comptonamplitude} \cite{BKKM,BKM}. Following the prescription
given in~\cite{babusci}, we subtract at this order the ``nucleon pole''
terms\footnote{\HBCPT is a non-relativistic framework with an expansion in
  $1/M$, where $M$ denotes the nucleon mass. The ``nucleon pole''
  contributions in the Mandelstam variables $s,\,u$ defined in \cite{babusci}
  therefore can show up as ordinary polynomials in the photon energy $\omega$
  in (\ref{eq:polestuff}). To the order in perturbation theory we are working
  here, the identification of the nucleon pole contributions in the full
  Compton amplitudes $A_1,\,A_2$ is trivial, as can be clearly seen from
  (\ref{eq:polestuff}). At higher orders in perturbation theory, matters can
  become more involved, see e.g. the discussion of the spin polarisability
  $\gamma_{M1}$ given in \cite{chiral2000}.}
\begin{eqnarray}
         A_1^\mathrm{pole}(\omega)=-\frac{e^2}{M}+\calO(1/M^3)&,&
         A_2^\mathrm{pole}(\omega)=\frac{e^2\omega}{M^2}+\calO(1/M^3)\;\;.
         \label{eq:polestuff}
\end{eqnarray}
As evident from Fig.~\ref{fig:comptonamplitude}, the power counting of \HBCPT
determines that the leading order contributions to the (dynamical)
polarisabilities of the nucleon are generated solely by the surrounding pion
cloud. The dynamical polarisabilities will be real functions of $\omega$ as
long as the photon energy stays below the pion production threshold. Above the
pion production threshold, the polarisabilities will be complex quantities.
Following the discussion on the origins of dispersive effects, we therefore
expect to see a cusp-like behaviour in the energy dependence as the photon
energy approaches $\mpi$. However, we focus the quantitative part of our
discussion on Compton scattering below $\mpi$ as we do not expect that the
leading order calculation is sufficient to reproduce the correct strength and
shape of the energy dependence near the cusp.

Calculating the amplitudes $A_1,\,A_2$ in the cm frame from the diagrams in
Fig.~\ref{fig:comptonamplitude} to leading one loop order, $\calO(p^3)$,
according to \cite{BKKM,BKM} and then subtracting the nucleon pole
contributions at this order with the help of (\ref{eq:polestuff}), one obtains
the Compton structure amplitudes $\bar{A}_1,\,\bar{A}_2$. Inserting them into
(\ref{eq:polarisabilitiesextr1}) yields the leading one loop, $\calO(p^3)$,
\emph{iso-scalar} dynamical electric and magnetic dipole and quadrupole
polarisabilities in \HBCPT:
\begin{eqnarray}
  \label{eq:HBCPTpols1}
  \alpha_{E1}^{(s)}(\omega)\Big|_{l=2}&=& 
  \frac{{\omega }^2}{12} \;\beta_{M2}(\omega)\;+
  \;\frac{e^2 g_A^2}{2\;(4\pi \fpi)^2}\;\frac{M}{W\omega}\;
  \int\limits_0^1\dd\!x\;\Bigg[\du -\sqrt{\du^2-1} + \nonumber\\
       &&+\;
  \frac{\left( 1 + \du^2 \right) }{\sqrt{2}}\;
     \left( \arctan[\frac{1}{\sqrt{2} \du}] -
       2 \arctan[\frac{1-x}{\sqrt{2} {\sqrt{\du^2 - x^2}}}
          ] \right) \Bigg] 
      \nonumber\\
  %%%%%%%%
  \beta_{M1}^{(s)}(\omega)\Big|_{l=2}&=&\frac{e^2 g_A^2}{16\;
  (4\pi\fpi)^2}
  \;\frac{M}{W \omega }\;\frac{1}{1 + 2 \du^2}\;
    \int\limits_0^1\dd\!x\;\Bigg[
    \sqrt{2}\left(2\du^4-\du^2-1\right)\arctan[\frac{1}{\sqrt{2}\du}] -
    \nonumber\\
       &&-
       2 \left( \du + \du^3 -2 \sqrt{2} \left( 1 + 2 \du^2 \right)  
          \arctan[\frac{1-x}{\sqrt{2}{\sqrt{\du^2 - x^2}}}]
         \right)  \Bigg]
      \nonumber\\
  %%%%%%%%
  \alpha_{E2}^{(s)}(\omega)\Big|_{l=2}&=&\frac{3 e^2 g_A^2}{4 \;
           (4\pi\fpi)^2}
           \;\frac{M}{W{\omega }^3}\;
    \int\limits_0^1\dd\!x\;\Bigg[ 
      \sqrt{2} 
          \left( \du^2-1 \right)  
            \arctan[\frac{1}{\sqrt{2} \du}] -\;\frac{2 \du 
         \left( 1 + \du^2 \right) }{1 + 
         2 \du^2 }\; - \nonumber\\
       &&- 
      \;\frac{8 \left( 1 + \du^2 \right)\left( x-1 \right)
        {\sqrt{\du^2 - x^2}}}{1 + 2 \du^2 -x \left( 2 + x \right)  }
        \; -4 \sqrt{2}\du^2\arctan[\frac{1-x}{\sqrt{2}{\sqrt{\du^2 - x^2}}}]
             \Bigg] 
      \nonumber\\
      %%%%%%%%%%%
   \beta_{M2}^{(s)}(\omega)\Big|_{l=3}&=&\frac{e^2 g_A^2}{8\;
      (4\pi \fpi)^2}\;  
    \frac{M}{W {\omega }^3}\;
    \int\limits_0^1\dd\!x\;\Bigg[
    -\;\frac{2 \du\left( 1 + 3 {\du}^2 +6 {\du}^4 \right)}{{\left( 1 + 
        2 {\du}^2 \right) }^2} \;+\; 
  \frac{8 \left(x-1 \right)  
     \sqrt{\du^2 - x^2}}{1 +  2 \du^2 - x \left( 2 + x \right) }\;+\nonumber\\
       &&\;+ \;\sqrt{2}\;\left(\left( 3 \du^2-1 \right)
         \arctan[\frac{1}{\sqrt{2} \du}] - 
     4 \arctan (\frac{x-1}
        {{\sqrt{2}} {\sqrt{{\du}^2 - x^2}}}) \right)
             \Bigg]  \nonumber\\
\end{eqnarray}
with $\fpi=92\;\MeV$ the pion decay constant and $\du:=\frac{\mpi}{\omega}$.
The integrals over the Feynman parameter $x$ can be performed by standard
methods, but the outcome is not especially enlightening and hence is omitted
here.

\begin{figure}[!htb]
  \begin{center}
    \includegraphics*[width=0.48\textwidth]{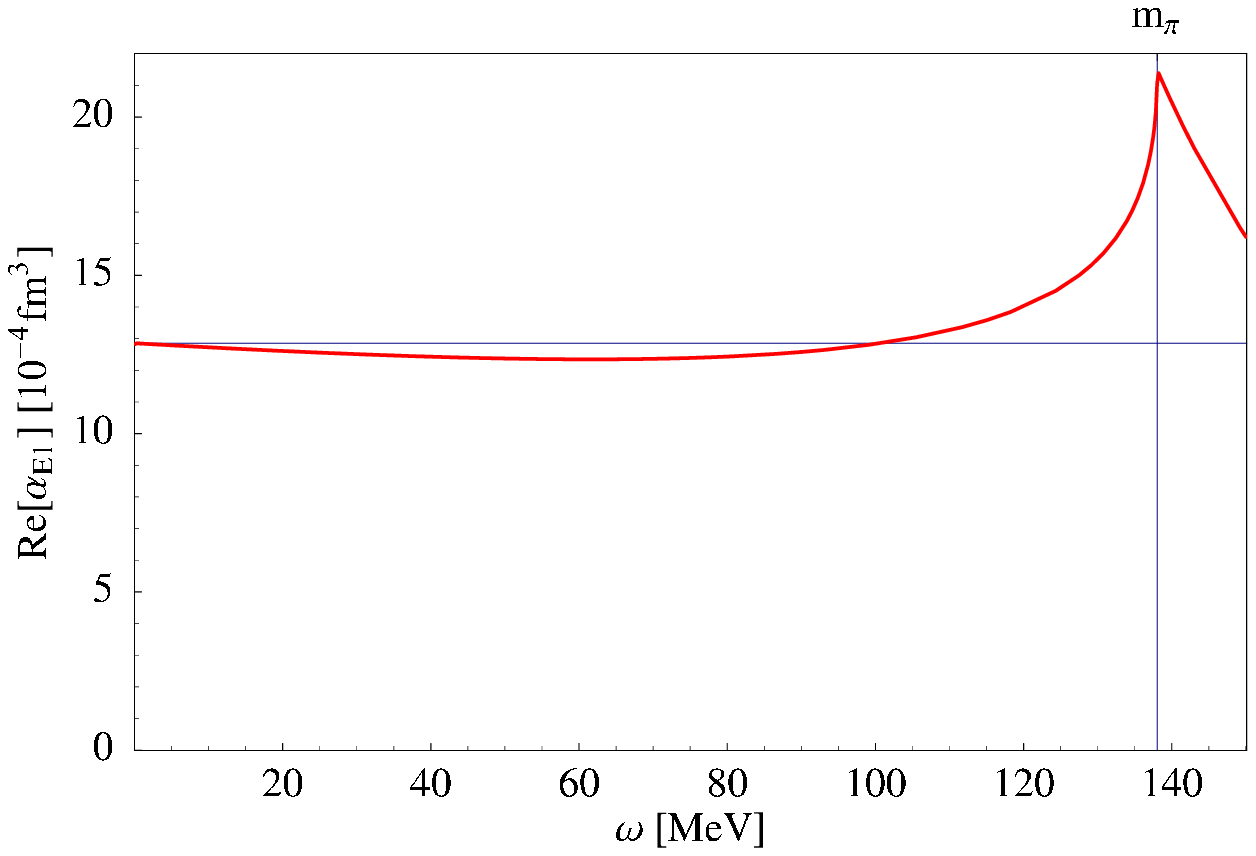}
    \hfill \includegraphics*[width=0.48\textwidth]{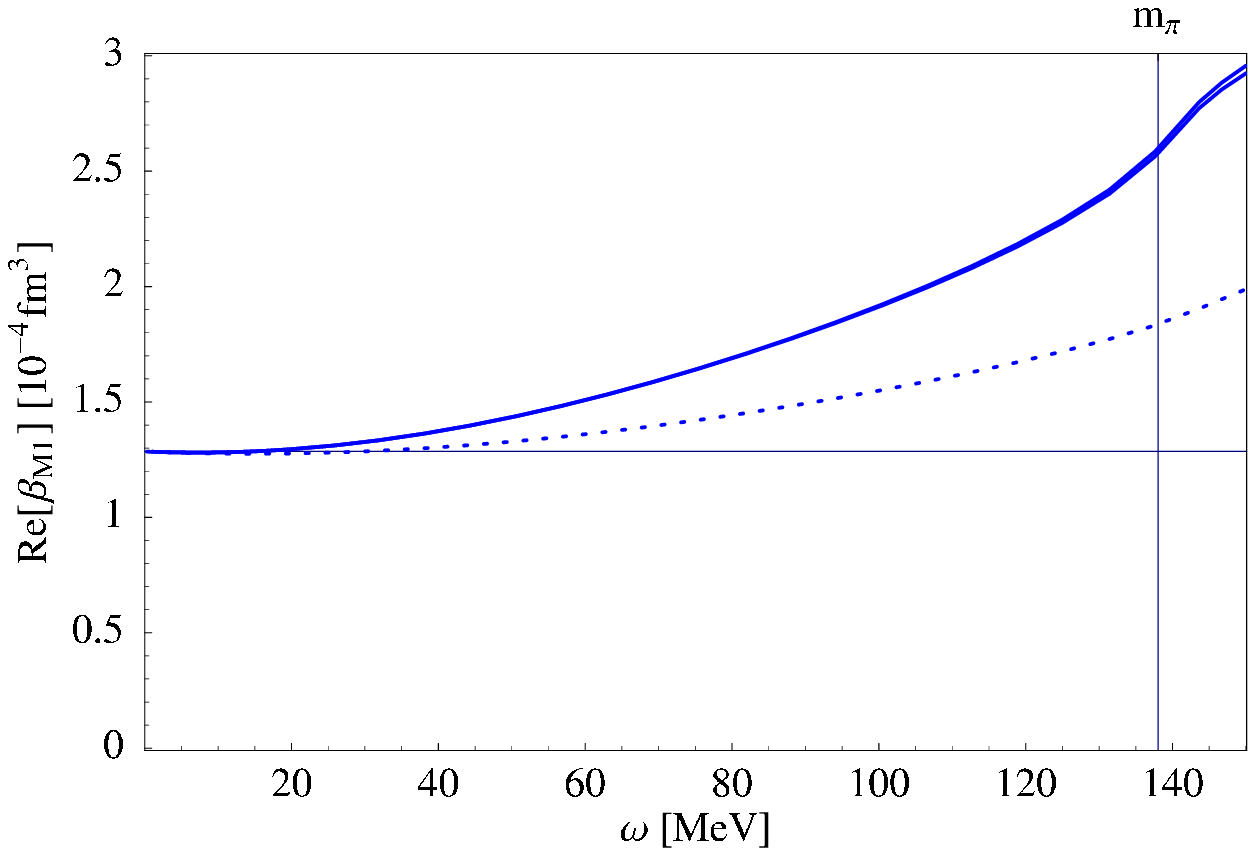}\\[2ex]
    \includegraphics*[width=0.48\textwidth]{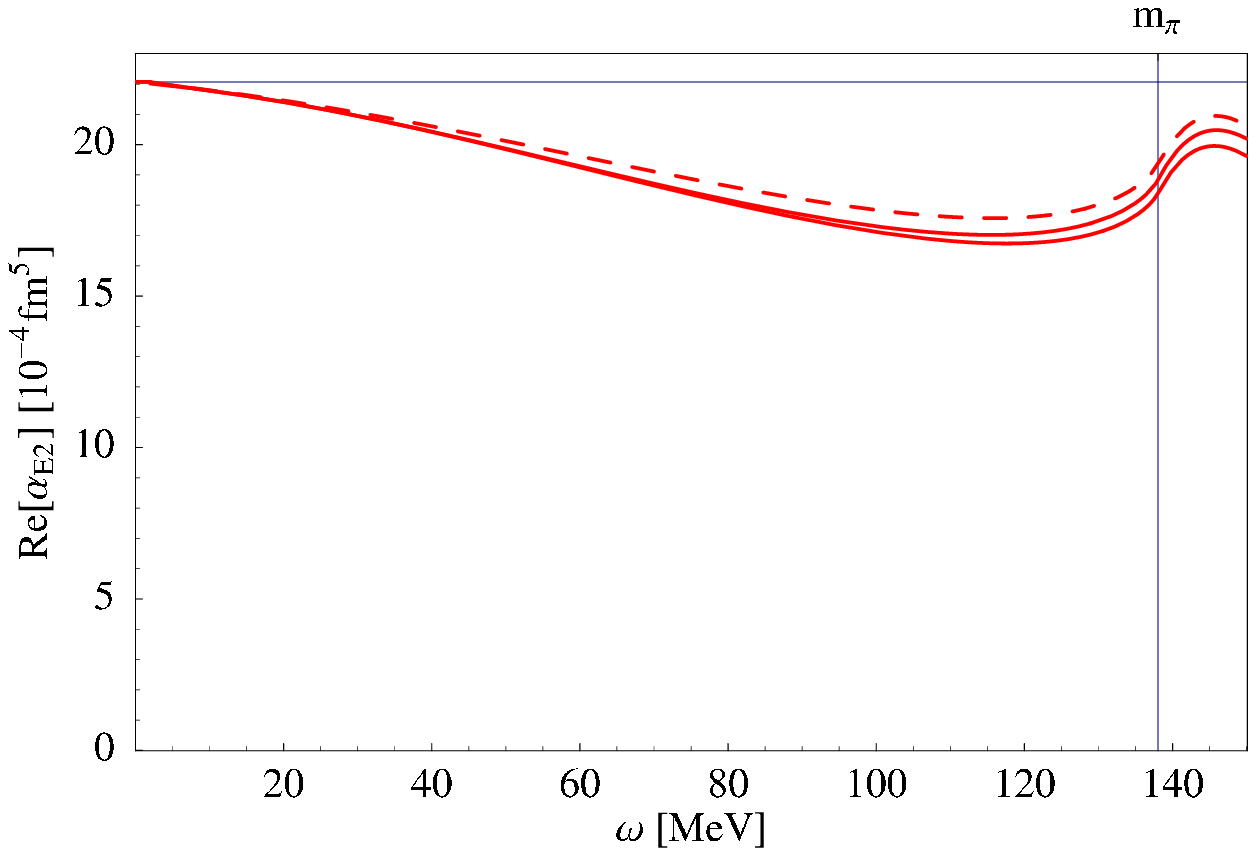} \hfill
    \includegraphics*[width=0.48\textwidth]{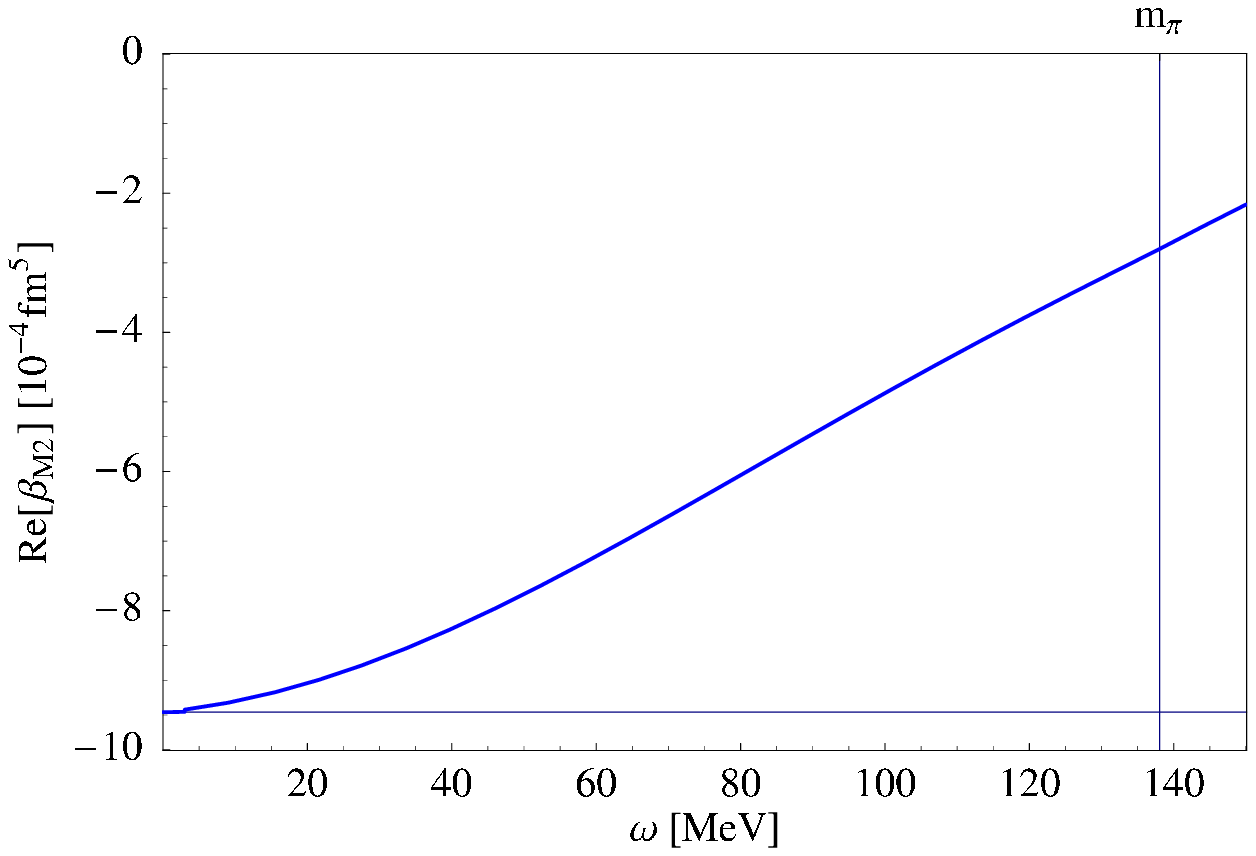}
    \caption{Leading one loop order HB$\chi$PT prediction for the dependence 
      of the dynamical \emph{iso-scalar} electric and magnetic dipole (top) and
      quadrupole (bottom) polarisabilities on the photon energy. Solid
      (dotted; dashed) lines: extraction truncated at $l=3$ ($l=1$; $l=2$).
      Notice the different scales. The upper left figure shows indeed three
      lines.}
    \label{fig:polarisabilities}
  \end{center}
\end{figure}

Recall that iso-scalar quantities are defined as the average of the
corresponding proton and neutron values. The (dynamical) \emph{iso-vector}
dipole and quadrupole polarisabilities are identically zero to this order in
the chiral calculation and hence are not considered. Notice that we give the
expressions for the polarisabilities $\alpha_{E1}^{(s)},\;\alpha_{E2}^{(s)}$
and $\beta_{M1}^{(s)}$ only up to $l=2$, while we choose the truncation at
$l=3$ for $\beta_{M2}^{(s)}$ to increase the numerical accuracy as discussed
below. To derive the complete result up to $l=3$ is straightforward, but we
refrain from quoting it as the expressions are lengthy and the numerical
difference between $l=2$ and $l=3$ for $\alpha_{E1}^{(s)},\;\alpha_{E2}^{(s)}$
and $\beta_{M1}^{(s)}$ is tiny. Instead, we plot the dynamical
polarisabilities of the $l=3$ truncation in Fig.~\ref{fig:polarisabilities}
together with results for $l=1$ and $l=2$ truncations, using the axial nucleon
coupling $g_A=1.27$ and the iso-scalar masses $\mpi=138\;\MeV$ and
$M=939\;\MeV$.

In the limit $\omega\rightarrow 0$, the expressions (\ref{eq:HBCPTpols1})
recover the well-known leading order \HBCPT predictions for the iso-scalar
electric and magnetic dipole polarisabilities
$\bar{\alpha}_E^{(s)},\,\bar{\beta}_M^{(s)}$ \cite{BKKM,BKM}, as well as the
predictions for their less known iso-scalar quadrupole counterparts
$\bar{\alpha}_{E2}^{(s)},\,\bar{\beta}_{M2}^{(s)}$ \cite{babusci}.  This can
easily be checked by Taylor expanding (\ref{eq:HBCPTpols1}) in the photon
energy $\omega$:
\begin{eqnarray}
  \label{eq:alphaslope}
  \alpha_{E1}^{(s)}(\omega\to0)
        &=&\frac{5 e^2 g_A^2}{24\;(4\pi\fpi)^2\;\mpi}\;
           \left[
             \left(1-\frac{\omega}{M}+\frac{\omega^2}{2M^2}
             \right)+\frac{9}{100}\;\frac{\omega^2}{\mpi^2}
             +\calO(\omega^3)\right]
           \nonumber\\
  \beta_{M1}^{(s)}(\omega\to0)
        &=&\frac{e^2 g_A^2}{48\;(4\pi\fpi)^2\;\mpi}\;
           \left[
             \left(1-\frac{\omega}{M}+\frac{\omega^2}{2M^2}
             \right)+\frac{7}{5}\;\frac{\omega^2}{\mpi^2}
             +\calO(\omega^3)\right]
           \nonumber\\
  \alpha_{E2}^{(s)}(\omega\to0)
        &=&\frac{7 e^2 g_A^2}{40\;(4\pi\fpi)^2\;\mpi^3}\;
           \left[
             \left(1-\frac{\omega}{M}+\frac{\omega^2}{2M^2}
             \right)-\frac{521}{1176}\;\frac{\omega^2}{\mpi^2}
             +\calO(\omega^3)
            \right]
           \nonumber\\
  \beta_{M2}^{(s)}(\omega\to0)
        &=&-\frac{3 e^2 g_A^2}{40\;(4\pi\fpi)^2\;\mpi^3}\;
           \left[
             \left(1-\frac{\omega}{M}+\frac{\omega^2}{2M^2}
             \right)-\frac{47}{42}\;\frac{\omega^2}{\mpi^2}
             +\calO(\omega^3)\right]
\end{eqnarray}
Note that the Taylor expansion in the photon energy at $\calO(p^3)$ does not
correspond to a chiral $1/M$ expansion, but that the intrinsic scale $\mpi$
connected to the structure of the nucleon enters as well. The terms in
(\ref{eq:alphaslope}) which explicitly depend on the nucleon mass stem from
expanding the kinematical factors $M/W$ in
(\ref{eq:polarisabilitiesextr1}/\ref{eq:polarisabilitiesextr2}) and are
expected to receive corrections from the next order, $\calO(p^4)$,
calculation.  We only show (\ref{eq:alphaslope}) to emphasise once more that
our dynamical polarisabilities are defined in the cm frame and therefore are
\emph{not} required to be symmetric functions in $\omega$.

The plots in Fig.~\ref{fig:polarisabilities} show that for the four leading
spin independent dynamical polarisabilities there is hardly any visible
difference between the $l=2$ and $l=3$ truncations of the Compton structure
amplitudes as defined in (\ref{eq:amplitudes1}), at least in the energy range
below the one pion threshold. The most pronounced effects related to the
truncation of the multipole series of the amplitude can be found in
$\alpha_{E2}^{(s)}$ for $\omega\sim110\;\MeV$, but even there the effect is on
the $5\%$ level.  Furthermore, the two formulae for the extraction of
$\beta_{M1}(\omega)$ and $\alpha_{E2}(\omega)$
(\ref{eq:polarisabilitiesextr1}/\ref{eq:polarisabilitiesextr2}) can be seen in
Fig.~\ref{fig:polarisabilities} to agree at the percent level, justifying the
omission of all $l\geq 4$ contributions in the multipole expansion of the
amplitudes. We also note that if one is only interested in dynamical dipole
polarisabilities, even the $l=2$ truncation seems to suffice. This means that
in $\bar{A}_1$, one only needs information up to quadratic terms in
$\cos\theta$, and in $\bar{A}_2$ even only the linear dependence is required
for a multipole projection.

We now turn to the physics discussion of each polarisability shown in
Fig.~\ref{fig:polarisabilities}.
\begin{itemize}
\item The surprising find -- at least to leading one loop order in \HBCPT --
  in the dynamical iso-scalar electric dipole polarisability
  $\alpha_{E1}^{(s)}$ is that out to photon energies of $100\;\MeV$, there is
  hardly any significant energy dependence. The contribution from the pion
  cloud produces a large value for $\bar{\alpha}_E^{(s)}$ which stays nearly
  constant throughout the low energy region. The only pronounced energy
  dependence results from the cusp-effect of the one pion production threshold
  and is expected following the discussion at the end of
  Sect.~\ref{sec:multipole}. Clearly, this finding has to be checked at the
  next order in \HBCPT~\cite{hgth}. If the pion cloud contributions then still
  do not result in a significant energy dependence apart from the expected
  cusp, any strong deviation from the
  $\alpha_{E1}^{(s)}(\omega)\approx\bar{\alpha}_E$ behaviour found from
  dispersion analyses could point to degrees of freedom in the nucleon beyond
  the usual pion nucleon dynamics. Details of this cusp structure are, of
  course, unlikely to be reproduced correctly by this low order calculation.
\item In contrast to $\alpha_{E1}^{(s)}$, the magnetic polarisabilities show
  large dispersive effects even at moderately low energies: At
  $\omega=100\;\MeV$, $\beta_{M1}^{(s)}$ is increased by $60\%$. Still, since
  it is small compared to $\alpha_{E1}^{(s)}$, this increase is small in
  absolute numbers.  The effect of the cusp is again clearly visible in the
  different slopes above and below the one pion production threshold.  At low
  energies, a relaxation mechanism is clearly at work.  We note that the
  energy dependence in $\beta_{M1}^{(s)}$ is interesting because with its help
  one might be able to identify the dynamical origin of the rather small
  overall value for the magnetic dipole polarisability of the proton
  (\ref{eq:globala}). More speculation on the dynamical content in
  $\beta_{M1}^{(s)}$ can be found in the concluding section of this
  presentation.
\item The dynamical iso-scalar magnetic quadrupole polarisability
  $\beta_{M2}^{(s)}$ perhaps shows the most surprising behaviour. At
  $\omega=100\;\MeV$, it drops to half its static value. Similar to a
  relaxation effect discussed in Sect.~\ref{sec:multipole}, it seems to loose
  most of its strength in the low energy region. Once more, we emphasise that
  the only dynamical degrees of freedom contained at this order are connected
  with the pion cloud of the nucleon. It will be interesting to see whether
  higher order chiral calculations or dispersion theory can substantiate this
  disappearance of magnetic quadrupole strength in the nucleon.
\item Finally, one observes at low energies a similar albeit less pronounced
  relaxation behaviour in the dynamical iso-scalar electric quadrupole
  polarisability $\alpha_{E2}^{(s)}$. However, around
  $\omega\approx100\;\MeV$, a second effect of opposite sign connected to the
  cusp of the one pion threshold is taking over, preventing a sharp falloff in
  the low energy region. Again, any energy dependence resulting from the
  electric quadrupole excitation of e.g.~the $\Delta$(1232) or other nucleon
  resonances as an intermediate state are not accounted for at this order in
  the calculation.
\end{itemize}   
We also note that although $\alpha_{E1}^{(s)}$ is hardly energy dependent at
low energies, the dispersive effects in $\beta_{M1}^{(s)}$ can lead to a
noticeable increase in the dynamical equivalent of the Baldin sum rule,
$\alpha_{E1}^{(s)}(\omega)+\beta_{M1}^{(s)}(\omega)$, from its iso-scalar
static value $13.2\times10^{-4}\;\fm^3$.

%%%%%%%%%%%%%%% Intro %%%%%%%%%%%%%%%%%%%
\section{Comments on Compton Scattering off the Deuteron}
\setcounter{equation}{0}
\label{sec:comments}
%%%%%%%%%%%%%%%%%%%%%%%%%%%%

Before concluding, we briefly dwell on low energy Compton scattering off the
deuteron. In contrast to low energy Compton scattering on the proton which is
not particularly sensitive to the polarisabilities due to at least equally
large effects from the magnetic moment terms, the effects from the nucleon
iso-scalar magnetic moments are small in the case of the deuteron, and the
sensitivity to polarisability effects is therefore greatly enhanced. As the
deuteron is \emph{modo grosso} an iso-scalar target, the iso-scalar
polarisabilities $\alpha_{El}^{(s)}$ and $\beta_{Ml}^{(s)}$ of the previous
section are accessed directly.

It has been noticed by various authors, e.g.~\cite{Kolb00,Levchuk01} and
\cite{Beane}, that the Compton scattering data on the deuteron taken by the
SAL group at $\omega_\mathrm{Lab}=95\;\MeV$~\cite{SAL} seem to prefer a static
iso-scalar magnetic dipole polarisability $\bar{\beta}_M^{(s)}\approx 7 \times
10^{-4}\;\fm^3$ which deviates strongly from dispersion analysis predictions
and from the proton values, while the static iso-scalar electric dipole
polarisability $\bar{\alpha}_E^{(s)}\approx 11\times 10^{-4}\;\fm^3$ is of the
expected size.  This leads also to a significant violation of the Baldin sum
rule, if the values extracted are taken as the \emph{static} dipole
polarisabilities. In view of the leading order \HBCPT result discussed in the
last section, one may speculate these findings to be a sign that the
dispersive effects in the iso-scalar single nucleon amplitudes might not be
under full control, as the leading order chiral calculation actually hints at
a basically unchanged $\alpha_{E1}^{(s)}(\omega)\approx\bar{\alpha}_E^{(s)}$
and a rise in $\beta_{M1}^{(s)}(\omega)$ with energy, albeit on a much smaller
scale. Partially, such dispersive effects have of course been taken into
account in~\cite{Kolb00,Levchuk01}. Likewise, Ref.~\cite{Beane} took the full
energy dependence of the leading one loop predictions for the \HBCPT
amplitudes $\bar{A}_1$ and $\bar{A}_2$ as input, together with its prediction
$\bar{\beta}_{M1}^{(s)}=1.2\times10^{-4}\;\fm^3$. None of these measures lead
to a resolution of the puzzle of the iso-scalar polarisabilities of the
nucleon. Clearly, our results given in the previous section cannot provide the
solution. However, we emphasise again that large dispersive effects in
$\beta_{M1}^{(s)}(\omega)$ could be at the heart of this problem\footnote{For
  example, the best microscopic calculation of the static proton
  polarisabilities $\bar{\alpha}_E^{(p)},\,\bar{\beta}_M^{(p)}$ in an
  effective field theory available today \cite{BKSM} attempts to model the
  finite parts of $\calO(p^4)$ counterterms via a resonance saturation
  hypothesis. It would be interesting to see what kind of energy dependence in
  $\beta_{M1}^{(p)}(\omega)$ results from the hypothesis of \cite{BKSM} that
  the smallness of $\bar{\beta}_M^{(p)}$ of the proton stems from a
  cancellation between a large paramagnetic effect arising from $\Delta(1232)$
  pole graphs and a large diamagnetism from the pion cloud. The implied energy
  dependence could be tested via the phenomenological results of dispersion
  analysis.} and we believe that the formalism outlined here allows for a
systematic study of these effects by comparing state of the art dispersion
analyses for nucleon Compton scattering with the best available microscopic
calculations of the relevant dynamics in the nucleon. We also note that it
might be interesting to re-analyse the SAL data by directly extracting the
values of the \emph{dynamical} polarisabilities following
(\ref{eq:polarisabilitiesextr1}/\ref{eq:polarisabilitiesextr2}) in a model
independent way. But before one can apply the formalism presented here at the
cross section level, one of course first has to provide a similar analysis for
the corresponding dynamical spin polarisabilities~\cite{hgth}.

Before concluding, we also want to comment on a determination of the
iso-scalar dipole polarisabilities of the nucleon from deuteron Compton
scattering at very low energies. It was suggested in Ref.~\cite{hggr} that a
window exists in the region $\omega\sim 20-50\;\MeV$ in which the nucleon
polarisabilities can be extracted in a model independent way without having to
know detailed pion dynamics. The authors present a feasibility study with data
at $\omega=48\;\MeV$.  Dispersive effects were estimated as inducing at most a
change on the level of $(\frac{\omega}{\mpi})^2\sim10\%$. The leading one loop
order \HBCPT analysis in the last section interestingly also shows that
$\alpha_{E1}^{(s)}$ and $\beta_{M1}^{(s)}$ are nearly energy independent in
that low energy r\'egime, with changes of less than $10\%$. We thus confirm
the authors' findings that at these low energies, the polarisabilities
extracted can be taken as the static ones, and that a two parameter fit of the
Compton scattering cross section to $\bar{\alpha}_E$ and $\bar{\beta}_M$ seems
justified.  Such a procedure also provides a valuable cross check, as the
Baldin sum rule is not used as input.

%%%%%%%%%%%%%%% Intro %%%%%%%%%%%%%%%%%%%
\section{Summary and Outlook}
\setcounter{equation}{0}
\label{sec:outlook}
%%%%%%%%%%%%%%%%%%%%%%%%%%%%

To conclude, we want to encourage the study of the \emph{dynamical} multipole
polarisabilities of the nucleon as an important tool to learn about its
internal structure. In contradistinction to the well known static
polarisabilities which measure the deformation of an object only in a static
electro-magnetic field, the dynamical polarisabilities encountered in many
areas of physics gauge the global response to an external, real photon of
arbitrary energy. Therefore, they contain additional information about the
dispersive effects, i.e.~the energy dependence of the polarisabilities, as
emphasised in Sect.~\ref{sec:multipole}. As they encode the excitation
spectrum of the internal degrees of freedom of a composite object in an
external electro-magnetic field of definite multipolarity, the nucleon
polarisabilities will necessarily be energy dependent due to relaxation
effects, resonances and particle production thresholds. Thus, properties of
the internal degrees of freedom of the nucleon are directly probed by them.

We defined the dynamical polarisabilities starting from a multipole projection
of the nucleon pole subtracted Compton scattering amplitudes and gave explicit
formulae how to extract the four leading spin independent multipole
polarisabilities (Sect.~\ref{sec:matching}). As a simple example, we
considered the predictions of leading one loop order \HBCPT for the dynamical
iso-scalar electric and magnetic dipole and quadrupole polarisabilities in
Sect.~\ref{sec:example}. We demonstrated that in this case, truncating the
non-pole Compton amplitudes at multipolarity $l=3$ is sufficient to obtain
numerically stable and convergent results for the dipole and quadrupole
polarisabilities. One of the important results of the chiral calculation is
that the energy dependence is likely to be strong for the magnetic
polarisabilities even far below the one pion production threshold. This might
have consequences for the recent discussions on Compton scattering off the
deuteron, as discussed in Sect.~\ref{sec:comments}.

The extension of the formalism to spin polarisabilities is
straightforward~\cite{hgth}.  In the future, we plan to identify the degrees
of freedom relevant for nucleon polarisabilities at low energies by comparing
the energy dependence for a set of dynamical nucleon polarisabilities,
e.g.~calculated in $\calO(p^4)$ $SU(2)$ \HBCPT, with the corresponding curves
from dispersion analyses~\cite{hgth}\footnote{One could imagine the following
  scenario: While the counterterms of \HBCPT (which parameterise all physics
  outside the pion nucleon particle space) can be fitted to the dispersion
  results obtained for the static polarisabilities at $\omega=0$, resulting
  discrepancies in the energy dependence in different multipole
  polarisabilities could point to the importance of explicit $\Delta$(1232),
  vector meson, kaon-cloud excitations etc.}. Finally, we hope that this
presentation has reminded the reader that nucleon Compton scattering can
provide much more dynamical information on the low energy structure of the
nucleon than is contained in the discussion of the usual, static dipole
polarisabilities.

%%%%%%%%%%%%%%%%%%%%%%%%%%%%%%%%%%%%%%%%%%%%%%%%%%%%%%%%%%%%%%%%%%%%%%%%%%%%%%%
%%%%%%%%%%%%%%%%%%%%%%%%%%%%%%%%%%%%%%%%%%%%%%%%%%%%%%%%%%%%%%%%%%%%%%%%%%%%%%%
%%%%%%%%%%%%%%%%%%%%%%%%%%%%%%%%%%%%%%%%%%%%%%%%%%%%%%%%%%%%%%%%%%%%%%%%%%%%%%%

\section*{Acknowledgements}

We are much indebted to the hospitality of the ECT* (Trento), and among its
members especially to B.~Pasquini and W.~Weise for fruitful discussions which
supported this work. In addition, the participants of the spring 2001
Collaboration Meeting on ``Real and Virtual Compton Scattering off the
Nucleon'' at the ECT* created a highly inspiring atmosphere. We also thank
N.~Kaiser, D.~R.~Phillips, G.~Rupak and S.~Scherer for stimulating discourse.
The authors are especially grateful to A.~L'vov for his input and his
clarifying comments regarding the conventions of Ref.~\cite{babusci}. We
acknowledge support in part by the DFG Sachbeihilfe GR 1887/1-2 (H.W.G.) and
by the Bundesministerium f{\"u}r Bildung und Forschung.

\newpage

%%%%%%%%%%%%%%%%%%%%%%%%%%%%%%%%%%%%%%%%%%%%%%%%%%%%%%%%%%%%%%%%%%%%%%%%%%%%%%%
%%%%%%%%%%%%%%%%%%%%%%%%%%%%%%%%%%%%%%%%%%%%%%%%%%%%%%%%%%%%%%%%%%%%%%%%%%%%%%%
%%%%%%%%%%%%%%%%%%%%%%%%%%%%%%%%%%%%%%%%%%%%%%%%%%%%%%%%%%%%%%%%%%%%%%%%%%%%%%%


\begin{thebibliography}{99}
  
\bibitem{babusci} D.~Babusci, G.~Giordano, A.~I.~L'vov, G.~Matone and
  A.~M.~Nathan: \journal{\PRC}{58}{1998}{1013} [hep-ph/9803347].
  
\bibitem{Olmos} V.~Olmos de Leon et al.: \journal{\EPJA}{10}{2001}{207}.
  
\bibitem{Kolb00} N.~R.~Kolb {\it et al.}: \journal{\PRL}{85}{2000}{1388}
  [nucl-ex/0003002].
    
\bibitem{Levchuk01} M.~I.~Levchuk and A.~I.~L'vov:
  \journal{\NPA}{684}{2001}{490} [nucl-th/0010059].
  
\bibitem{hggr} H.~W.~Grie\3hammer and G.~Rupak: \PLB\ in print
  [nucl-th/0012096].
  
\bibitem{Galler} G.~Galler et al.: \journal{\PLB}{503}{2001}{245}
  [nucl-ex/0102003].
  
\bibitem{Blanpied} G.~Blanpied et al.: \journal{\PRC}{64}{2001}{025203}.
  
\bibitem{lvov} A.~I.~L'vov, V.~A.~Petrunkin and M.~Schumacher:
  \journal{\PRC}{55}{1997}{359}.
  
\bibitem{drechsel} D.~Drechsel, M.~Gorchtein, B.~Pasquini and
  M.~Vanderhaeghen: \journal{\PRC}{61}{2000}{015204} [hep-ph/0103172].
  
\bibitem{spin} see e.g.~T.~R.~Hemmert, B.~R.~Holstein, J.~Kambor and G.~Kn{\"
    o}chlein: \journal{\PRD}{57}{1998}{5746} [nucl-th/9709063] and references
  therein.
  
\bibitem{hgth} H.~W.~Grie\3hammer, T.~R.~Hemmert, R.~Hildebrandt and
  B.~Pasquinig: in preparation.
  
\bibitem{Ritus} V.~I.~Ritus: \textit{Sov.~Phys.~JETP} \textbf{5}, 1249 (1957).
  
\bibitem{Ritus2} A.~P.~Contogouris: \textit{Nuovo Cimento} \textbf{25}, 104
  (1962).
  
\bibitem{Ritus3} Y.~Nagashima: \textit{Prog.\ Theor.\ Phys.\ }\textbf{33}, 828
  (1965).
  
\bibitem{Gui} I.~Guia\c{s}u and E.~E.~Radescu: \textit{Ann.~Phys.\ 
    }textbf{120}, 145 (1979).

\bibitem{Guichon} P.~A.~M. Guichon, G.~Q. Liu and A.~W. Thomas:
  \journal{\NPA}{591}{1995}{606} [nucl-th/9605031].
  
\bibitem{L'vov:2001fz} A.~I.~L'vov, S.~Scherer, B.~Pasquini, C.~Unkmeir and
  D.~Drechsel: \journal{\PRC}{64}{2001}{015203} [hep-ph/0103172].
  
\bibitem{sixties} H.~Arenhovel and W.~Greiner:
  \textit{Prog.~Nucl.~Phys.}~\textbf{10}, 167 (1969). 
  
\bibitem{Pfeil} W.~Pfeil, H.~Rollnik and S.~Stankowski:
  \journal{\NPB}{73}{1974}{166}.
  
\bibitem{BKKM} V.~Bernard, J.~Kambor, N.~Kaiser and U.-G.~Mei{\ss}ner:
  \journal{\NPB}{388}{1992}{315}.
  
\bibitem{BKM} V.~Bernard, N.~Kaiser and U.-G.~Mei{\ss}ner: \textit{Int.\ J.\ 
    Mod.\ Phys.\ }\textbf{E4}, 193 (1995).
  
\bibitem{chiral2000} T.~R.~Hemmert in ``Chiral Dynamics 2000: Theory and
  Experiment III'', eds.~A.~M.~Bernstein, J.~L.~Goity and U.-G.~Mei\3ner,
  World Scientific 2001, [nucl-th/0101054].
  
\bibitem{Beane} S.~R.~Beane in ``Chiral Dynamics 2000: Theory and Experiment
  III'', eds.~A.~M.~Bernstein, J.~L.~Goity and U.-G.~Mei\3ner, World
  Scientific 2001, [nucl-th/0012042].
  
\bibitem{SAL} D.~L.~Hornidge et al.: \journal{\PRL}{84}{2000}{2334}
  [nucl-ex/9909015].
  
\bibitem{BKSM} V.~Bernard, N.~Kaiser, U.-G.~Mei{\ss}ner and A.~Schmitt:
  \journal{\PLB}{319}{1993}{269} [hep-ph/9309211].
  
  
  
\end{thebibliography}
\end{document}